# Critical comments on EEG sensor space dynamical connectivity analysis

Frederik Van de Steen[1], Luca Faes[2], Esin Karahan[3], Jitkomut Songsiri[4], Pedro A. Valdes-Sosa[3,5], Daniele Marinazzo[1]

[1]Department of Data Analysis, Ghent University, 9000 Ghent

[2]Healthcare Research and Innovation Program, FBK, Trento and
BIOTech, Dept. of Industrial, Engineering, University of Trento, 38123 Mattarello, Trento, Italy

[3]Key Laboratory for Neuroinformation of the Ministry of Education, UESTC, 610054 Chengdu, China

[4]Control Systems Laboratory, Electrical Engineering Department, Chulalongkorn University, 10330 Bangkok, Thailand

[5]Cuban Neuroscience Center, 15202 La Habana, Cuba

Corresponding Author: Frederik Van de Steen

E-mail:Frederik.vandesteen@ugent.be

Tel: 003292646482

## Acknowledgements

This research was supported by the Fund for Scientific Research-Flanders (FWO-V), grant FWO14/ASP/255.

## ABSTRACT

Many different analysis techniques have been developed and applied to EEG recordings that allow one to investigate how different brain areas interact. One particular class of methods, based on the linear parametric representation of multiple interacting time series, is widely used to study causal connectivity in the brain. However, the results obtained by these methods should be interpreted with great care. The goal of this paper is to show, both theoretically and using simulations, that results obtained by applying causal connectivity measures on the sensor (scalp) time series do not allow interpretation in terms of interacting brain sources. This is because 1) the channel locations cannot be seen as an approximation of a source's anatomical location and 2) spurious connectivity can occur between sensors. Although many measures of causal connectivity derived from EEG sensor time series are affected by the latter, here we will focus on the well-known time domain index of Granger causality (GC) and on the frequency domain directed transfer function (DTF). Using the state-space framework and designing two simulation studies we show that mixing effects caused by volume conduction can lead to spurious connections, detected either by time domain GC or by DTF. Therefore, GC/DTF causal connectivity measures should be computed at the source level, or derived within analysis frameworks that model the effects of volume conduction. Since mixing effects can also occur in the source space, it is advised to combine source space analysis with connectivity measures that are robust to mixing.

**Keywords:**

EEG, Brain connectivity, MVAR, Granger causality, Directed transfer function

## 1. Introduction

With the advent of non-invasive techniques such as, positron emission tomography (PET), functional magnetic resonance imaging (fMRI), electroencephalograph (EEG), magnetoencephalography (MEG) and more recently functional near-infrared spectroscopy (fNIRS), the study of the dynamical behaviour of the human brain has become a major field in science. Along with these techniques, many data analysis tools have been developed that allow researchers to investigate brain functioning to a greater extent. For many years, researchers were mostly interested in the localization of function both in space (e.g. fMRI) and time (e.g. EEG). More recently there has been a growing interest in functional integration, which refers to the interplay between functionally segregated brain areas (Friston 2011). This tendency has led to the development, refinement and use of advanced data analysis techniques such as dynamic causal modelling (DCM; Friston et al. 2003), independent component analysis (ICA, e.g. Makeig et al. 1996; Van De Ven et al. 2004), and methods based on multivariate autoregressive (MVAR) modeling. While each of these approaches has its own merits and pitfalls, here we will focus on the application of MVAR-based connectivity analysis on EEG data.

Mainly due to its close relation with the ubiquitous concept of Granger causality (GC, Granger 1969), the use of MVAR-based measures to investigate human brain connectivity has become increasingly popular in neuroscience (e.g. Ding et al. 2006; Liao et al. 2011; Porta and Faes, 2016; Seth et al. 2015). MVAR models are being extensively applied to quantify –in both time and frequency domains– the concept of GC expressing the direct causal influence from one series to another ( Baccala et al. 2001; Ding et al. 2006; Seth et al. 2015), as well as the related concept of total causal influence between two time series –expressed in the frequency domain by the so-called directed transfer function (DTF; Kamiński and Blinowska 1991). Although the use of MVAR modelling on neural time series has been debated (e.g. Friston et al. 2013; Friston et al. 2014; Seth et al. 2015), this approach is widely used in neuroscience due to its high flexibility (no assumptions are made about the underlying neural mechanisms) and easiness of implementation (the derived measures like GC and DTF are obtained from the model parameters estimated by standard least squares techniques). Moreover, there is a continuous refinement of MVAR-based methods in order to deal with issues related to applicative specific contexts. Recently, for example, linear state space (SS) models have been proposed for GC analysis, which can overcome the bias due to e.g. sampling and filtering (Barnett and Seth 2015), or account for the well know effect of volume conduction typical of EEG analysis (Cheung et al. 2010).

In this paper, we do not intend to debate the use of MVAR-based and GC analyses in neuroscience. Our goal is to show that MVAR-based analysis of causal connectivity performed at the level of EEG sensors does not allow interpretation in terms of anatomically interacting sources. There are two reasons why this interpretation is problematic. First, no inferences can be made about the anatomical location of the (possibly) interacting sources based on the location of the sensors (Nunez and Srinivasan 2006; Haufe et al. 2013; Papadopoulou et al. 2015). Secondly, since MVAR-based measures are influenced by volume conduction, their straight application to sensor EEG data can lead to the detection of spurious connections. The former problem is generally accepted in the literature. Regarding the latter, there is less consensus. This issue will be the main focus of the present work which mainly disagrees with what said in Kaminski and Blinowska (2014). In that paper the authors claim that MVAR-based measures such as the DTF are not influenced by volume conduction. Their main argument is that volume conduction is an instantaneous effect and hence cannot induce a phase shift at the sensors. Along that line on the other hand, several papers were published where the DTF was applied to the EEG sensor data to make claims regarding interacting brain sources (e.g. Wyczesany et al. 2014; Ligeza et al. 2015; Wyczesany et al. 2015). On the other hand, several authors have pointed to the adverse effects of volume conduction on sensor space connectivity analysis (Gómez-Herrero et al. 2008; Schoffelen and Gross 2009; Haufe et al. 2010; Haufe et al. 2013; Bastos and Schoffelen 2016). Vinck et al. (2015) for example showed that linearly mixed additive noise (as a consequence of e.g. volume conduction) can lead to spurious connectivity. Here we will investigate the effect of volume conduction on two of the most popular MVAR measures of causal connectivity: the time domain GC index (Geweke 1982) and the DTF (Kamiński and Blinowska 1991).

The paper is organized as follows. First, we show how brain dynamics can be modelled at the source level using MVAR models and how source activity is projected to the EEG channels (section 2.1). Then, we discuss how these dynamics can be quantified both in the time and frequency domain using GC and DTF measures (section 2.2). In section 2.3, we outline three strategies for causal connectivity analysis that may be applied to EEG data: 1) the analysis performed directly on sensor data; 2) a two stage approach where the source time series are first estimated after which the analysis is performed on these reconstructed time series; 3) the analysis performed within the state space (SS) framework. Then, we show how the first strategy can lead to spurious connectivity results (section 3). This is shown first by making some theoretical considerations regarding the effects of volume conduction on causal connectivity analysis of EEG sensor time series (section 3.1), and then illustrating the issue with two simulations studies (section 3.2). In the first simulation, the adverse effect of mixing are demonstrated with a simple 'toy' model. In the second simulation study, more realistic EEG data are generated. Finally, we discuss our findings in light of the existing literature.

## 2. Causal Connectivity Analysis of EEG Time Series

### *2.1 Linear Modelling of EEG Source Interaction and Volume Conduction*

The general aim of the application to neural data of methods for connectivity analysis like GC and DTF is to investigate how different functionally segregated brain areas interact with each other. In the case of the EEG, the data consists of electrical potentials which are measured using multiple electrodes that are placed at different locations of the scalp. These potentials are generally believed to originate from the summation of post-synaptic current flows that are generated by local synchronized activity of many neurons. These locally synchronized neurons are conveniently represented as a cortical dipole which we call a source (Baillet et al. 2001). Due to the propagation effect known as volume conduction, these source activations are instantly mixed resulting in electrical potentials that are measured at the scalp. In this study we represent the time series of $M$ cortical sources as realizations of the stochastic process $S(t) = [S_1(t) \cdots S_M(t)]^T$s, and the corresponding measured time series of $L$ scalp sensors as realizations of the process $Y(t) = [Y_1(t) \cdots Y_L(t)]^T$ (where $t$ = 1,...,$T$ and $T$ is the length of the realization). The dynamical interactions between the sources can be modelled by means of a MVAR model (Lütkepohl 2005)

$$S(t) = \sum_{d=1}^{p} \mathbf{A}(d) S(t-d) + U(t) \qquad (1)$$

where $\mathbf{A}(d)$, $d$=1,...,$p$, are $M$ x $M$ coefficient matrices, $p$ is the maximum time lag that is used for quantifying the influences of past states on the current state and $U(t)$ is a white noise innovation process with diagonal covariance matrix $\Theta = E[U(t)U(t)^T]$. The mapping from source space to sensor space can be expressed by (Haufe et al. 2013)

$$Y(t) = \mathbf{L}S(t) + G(t) \qquad (2)$$

where $G(t) = [G_1(t) \cdots G_L(t)]^T$ is the $L$-dimensional vector of the measurement noise superimposed to each channel and $\mathbf{L}$ is a $L$ x $M$ lead field matrix. The columns of the lead field matrix summarize the so-called forward model for the different sources. Due to volume conduction, the sources are mixed (more specifically, linearly superimposed) and hence the lead field matrix is non-sparse so that each source contributes to some extent to the measured scalp potentials. Assuming that different sources are activated, the EEG channels therefore contain information from a mixture of sources.

### *2.2 Causal Connectivity Measures*

Granger causality (GC) is a useful tool to quantify directed neural interactions in neuroscience (Porta and Faes 2016). The general principle of GC is based on predictability and temporal precedence: given two time series $S_1$ and $S_2$, respectively considered as the target and the driver, $S_2$ is said to Granger cause $S_1$ if the prediction of its current value is improved by considering the past values of $S_2$ compared to the case in which only the past value of $S_1$ are exploited for prediction (Granger 1969). When $M$>2 time series are considered, a distinction must be made between "direct" and "indirect" GC, where the former is obtained by conditioning on all the $M$-2 times-series other than the driver and the target, i.e., including all these other time series in the prediction models. The classical definition of "direct" GC from a driver series $S_j$ to a target series $S_i$ ($1 \leq i,j \leq M$) is based on linear MVAR models in the form of (1), where two models are formulated and predictability is assessed in terms of prediction error variance: the first model (i.e the "full" model) contains all time series exactly as in (1), and the predictability of the target series $S_i$ is assessed through the variance of $U_i$, $\Theta_{ii}^{F} = E[U(t)_i^2]$; the second model (i.e., the "reduced" model) contains all time series except the driver $S_j$, so that the predictability of $S_i$ is now assessed from the reduced innovations $U(t)_i^R$ as $\Theta_{ii}^{R} = E[{U(t)_i^R}^2]$. Then, time domain GC (TGC) from $S_j$ to $S_i$ is quantified in its logarithmic formulation (Geweke 1982) as

$$TGC_{ij} = \ln \frac{\Theta_{ii}^{R}}{\Theta_{ii}^{F}} \tag{3}$$

The superscripts $F$ and $R$ refer to the full and reduced MVAR model respectively. From (3), it is clear that if the inclusion of the driver does not reduce the variance of the error (i.e. the denominator) of the target compared to the model without the driver (i.e. the numerator), then the ratio of the error variances will be 1 and hence GC will be zero. If the variance in the error is reduced, the ratio will be bigger than one and hence the GC values will be larger than 0. This implies that at least one of the coefficients in the **A** matrices that relates the past values of the driver to the current value of the target is different from zero (Haufe et al. 2013).

Frequency domain measures of causal connectivity can be obtained by taking the Fourier transform of (1), yielding

$$S(\omega) = \sum_{d=1}^{P} \mathbf{A}(d)\, S(\omega) e^{-i2\pi dT\omega} + U(\omega) \tag{4a}$$

$$S(\omega) = \mathbf{H}(\omega) U(\omega), \mathbf{H}(\omega) = (\mathbf{I}_M - \sum_{d=1}^{P} \mathbf{A}(d)\, e^{-i2\pi dT\omega})^{-1} \tag{4b}$$

where $\mathbf{I}_M$ is the $M\times M$ identity matrix, $i = \sqrt{-1}$, and $\mathbf{H}(\omega)$ is the transfer function acting as a linear filter which relates the zero mean white noise processes $U(\omega)$ to the observed processes $S(\omega)$. More specifically, the element $H_{ij}(\omega)$ is the frequency response of $S_i$ to a zero phase unit amplitude sinusoidal chock with frequency $\omega$ in $S_j$. In the case of DTF, the elements of the transfer function are used to quantify in normalized terms the influence of the $j$-th innovation process on the $i$-th observed process, resulting in the definition (Kamiński et al. 2001)

$$\gamma_{ij}(\omega)^2 = \frac{\left|H_{ij}(\omega)\right|^2}{\sum_{m=1}^{K}\left|H_{im}(\omega)\right|^2} \tag{5}$$

The DTF is classically adopted to infer in the frequency domain the total directed influence from $S_j$ to $S_i$, i.e. the directed influence arising from both the direct and indirect causal connections between the two time series (Eichler 2006; Eichler 2012; Faes 2014). Importantly, since the DTF is based on the elements of the transfer function, one should keep in mind that the DTF measures the frequency response of one time series to a random sinusoidal chock (i.e. noise input) at another time series and not the frequency response to the time series itself. The interpretation of the DTF deviates from the GC notion and should be treated as a complementary measure of causal connectivity. If there are no interactions among the time-series and when each time-series only depends on its own past, the transfer function is a diagonal matrix and the DTF will be uniformly zero over frequencies for each time-series pair. If, on the contrary, there is at least one interaction among a pair of the time-series, then at least one of the off-diagonal elements in one of the $\mathbf{A}(d)$ matrices is non-zero and hence the transfer function $\boldsymbol{H}(\omega)$ will be a non-diagonal matrix, leading to a DTF which is not uniformly zero over the frequency range for at least one pair of channels.

*2.3 Assessment of Causal Connectivity for EEG time series*

In the assessment of directional coupling among different neural sources by means of GC and DTF analyses, it is clear that the goal is to make inferences on the interactions between the source time series $S_t$. A major issue in this case is that the source time series are not directly observable, and therefore the methods presented above cannot be applied in a straightforward way. Three possible approaches can be followed in this case, as discussed in the following.

The first approach is to perform causal connectivity analysis as described in Sect. 2.2 directly on the available sensor time series $Y(t)$ (Kami□ski a□d Bli□owska 2014; Wyczesa□y et al. 2014; Wyczesa□y et al. 2015). This implies considering the MVAR model

$$Y(t) = \sum_{d=1}^{p} \widetilde{\boldsymbol{A}}(d)Y(t-d) + \widetilde{U}_t(t), \quad (6)$$

which, being estimated from different time series than the sources $S(t)$, will in general yield MVAR parameters ($\widetilde{\boldsymbol{A}}(d), \widetilde{\Theta} = E[\widetilde{U}(t)\widetilde{U}(t)^T]$) that differ from those relevant to the interacting ( $\mathbf{A}(d)$, $\Theta$), sources and thus will lead to different values of GC and DTF.

The second way to proceed is to employ a two-stage approach, by (i) reconstructing the time course of the neural sources by means of some source reconstruction method (e.g. Michel et al. 2004) to get the time series $\hat{S}(t) = [\hat{S}_1(t) \cdots \hat{S}_M(t)]^T$ which represent an estimate of the unmeasured sources $S(t)$; and (ii) performing causal connectivity analysis on the reconstructed sources after identification of the MVAR model:

$$\hat{S}(t) = \sum_{d=1}^{p} \widehat{\boldsymbol{A}}(d)\hat{S}(t) + \widehat{U}(t). \quad (7)$$

In this case it is expected that, if the reconstruction method and the MVAR estimation approach are accurate enough, the MVAR parameters ($\widehat{\boldsymbol{A}}(d)$, $\widehat{\Theta}$) will be a good estimate of the true parameters ( $\mathbf{A}(d)$, $\Theta$), and so will be the corresponding GC and DTF values.

A third possibility is to exploit the framework of state-space (SS) models to combine source interaction and volume conduction models and estimate the measures of causal connectivity within this framework (Cheung et al. 2010; Barnett and Seth 2015). The general linear time-invariant SS model can be formulated as

$$X(t+1) = \mathbf{B}X(t) + W(t) \quad (8a)$$

$$Y(t) = \mathbf{C}X(t) + V(t) \quad (8b)$$

where $X(t)$ is an $K$-dimensional unobserved state process, $Y(t)$ the $L$-dimensional observed process, the $K \times K$ state transition matrix **B** describes the update of the hidden states at each discrete time increment, and the $L$ x $K$ observation matrix **C** describes the instantaneous mapping of the state process $X$ to the observed process $Y$; $W(t)$ (state-equation error) and $V(t)$ ( measurement equation error) are zero-mean white noise processes with covariances $\Phi = E[W(t)W(t)^T]$ and $\Omega = E[V(t)V(t)^T]$, and cross-covariance $\Sigma = E[W(t)V(t)^T]$. The source interaction and volume conduction models of Eqs. (1,2) can be re-written in SS form by defining the state process $X_t = [S(t)^T \cdots S(t-p)^T]^T$ of dimension $K$=$Mp$, defining the state and observation errors $W(t) = [U(t+1)^T 0 \cdots 0]^T$ and $V(t) = G(t)$, letting the state and measurement errors be uncorrelated ($\Sigma = 0$), and setting the parameters of the state and observation equations (8a,b) as:

$$\mathbf{B} = \begin{bmatrix} \mathbf{A}(1) & \cdots & \mathbf{A}(p-1) & \mathbf{A}(p) \\ \mathbf{I}_M & \cdots & \mathbf{0}_M & \mathbf{0}_M \\ \vdots & \ddots & \vdots & \vdots \\ \mathbf{0}_M & \cdots & \mathbf{I}_M & \mathbf{0}_M \end{bmatrix}, \qquad (9)$$
$$\mathbf{C} = [\mathbf{L} \quad \mathbf{0}_{L\times(K-M)}] \, .$$

An important property of the SS representation is that it allows to compute the time domain measure of GC directly from the model parameters, without the need of identifying the reduced model and estimate its prediction error variance (Barnett and Seth 2015). Indeed, time domain GC among the sources is assessed in a straightforward way deriving the parameters ($\mathbf{A}_d$, $\Theta$) from $\mathbf{B}$ and $\Omega$; GC among the sensors is assessed in the time domain from the “innovations form” representation of the SS model. In the frequency domain, the DTF is computed from the moving-average representation of the SS model, which can be easily derived from the original MVAR parameters (see, e.g., Barnett and Seth 2015, for a detailed description of the procedure).

## 3. Effect of Volume Conduction on Sensor-Space Causal Connectivity

### *3.1 Theoretical Considerations*

The easiest way to accomplish causal connectivity analysis on EEG datasets is to follow the first approach described in Sect. 2.3 (Eq. (6)), since it only requires the application of MVAR modelling routines on the observed sensor time series. However, in this case simplicity comes at a price. First of all, if one finds significant GC or DTF among pairs of sensors, one cannot make inferences about the spatial location of the underlying sources except for the (rare) case when one knows (or assumes) that the activity of one single source is expressed at only one EEG channel and that one knows where these source are a priori (Papadopoulou et al. 2015). If one cannot make these assumptions, the channel time series are a mixture of activated sources and therefore it is impossible to infer which source's past values that are expressed in the assigned driver sensor contribute to the prediction of another source's future which is expressed in the target sensor. Even when there are no interacting sources, a significant interaction between two sensors can be found applying standard connectivity analysis either in the time domain (e.g., the TGC index (3)) or in the frequency domain (e.g., the DTF (5)). This latter point has been recently debated (Kaminski and Blinowska 2014). In that paper, the authors claim the DTF is not influenced by volume conduction. If this would be true, then non-interacting VAR sources should not lead to a significant DTF at some frequencies between any pair of sensors. However, if the sources can be expressed by (1), combined with (2) to obtain the scalp potentials, we obtain the following relations for the time and frequency domain representations of the sensor series:

$$Y(t) = \mathbf{L}(\sum_{d=1}^{p} \mathbf{A}(d)S(\mathrm{t}-\mathrm{d}) + U(t)) + G(t) \tag{10a}$$

$$Y(\omega) = \mathbf{LH}(\omega)U(\omega) + \mathrm{G}(\omega) \tag{10b}$$

In case of non-interacting MVAR sources, the coefficient matrix **A** and the transfer function matrix **H**(ω) are both diagonal matrices. However, given that the lead field matrix **L** is non-diagonal, the products $\mathbf{LA}(d)$ in (10a) and $\mathbf{LH}$(ω) in (10b) result in a non-diagonal matrix. This implies that the sensors can be expressed as a weighted sum of past values of all the sources (though in practice the contribution of some specific sources to a specific EEG channel can be negligible). In addition, the mixing of the sources is different for each sensor. This implies that the past values of the sensors will improve the prediction of another sensor current state (even though the mixing is instantaneous, past values are also instantaneous mixtures of sources) and hence, spurious causal connectivity will occur both when measured by the TGC (non-diagonal coefficient matrix) and when measured by the DTF (non-diagonal transfer function).

The above considerations show theoretically that non-interacting MVAR sources may, in general, yield causal connectivity among sensors. In the following we identify some particular conditions for which the mixing of non-interacting sources yields zero causal connectivity between sensors. The first case is that of mixing of pure random statistically independent noise time series, for which the sensor series are $S(t) = U(t)$ and the sensor series become $Y(t) = LU(t) + G(t)$, thus exhibiting zero-lag correlation but no temporal structure, and thus no time-lagged correlations. Another situation is that of non-interacting MVAR sources with proportional power spectra projected without measurement noise. In this case the sources obey to Eq. (1) with $\mathbf{A}(d) = a_d\mathbf{I}_M$, where $a_d$ are constant scalars and $\mathbf{I}_M$ is the $M\times M$ identity matrix, and the mapping occurs according to Eq. (2) with $G(t)=0$. Under these assumptions, Eq. (10a) can be equated to Eq. (6) yielding $\widetilde{U}(t) = \mathbf{L}U(t)$ and $\widetilde{\mathbf{A}}(d) = \mathbf{LA}(d)\mathbf{L}^{+}$, where $\mathbf{L}^{+}$ is the Moore-Penrose pseudoinverse of **L**. Then, we distinguish three cases: if **L** is square (there are as many sources as sensors, $L=M$), we have $\widetilde{\mathbf{A}}(d) = \mathbf{LA}(d)\mathbf{L}^{-\mathbf{1}} = a_d\mathbf{I}_L$; if **L** is fat (there are more sources than sensors, $L<M$), we have $\widetilde{\mathbf{A}}(d) = \mathbf{LA}(d)\mathbf{L}^{\mathrm{T}}(\mathbf{LL}^{\mathrm{T}})^{-1} = \mathbf{L}a_d\mathbf{I}_M\mathbf{L}^{\mathrm{T}}(\mathbf{LL}^{\mathrm{T}})^{-1} = a_d\mathbf{I}_L$; if **L** is skinny (there are more sensors than sources, $L>M$), we have $\widetilde{\mathbf{A}}(d) = \mathbf{LA}(d)(\mathbf{L}^{\mathrm{T}}\mathbf{L})^{-1}\mathbf{L}^{\mathrm{T}} = a_d\mathbf{L}(\mathbf{L}^{\mathrm{T}}\mathbf{L})^{-1}\mathbf{L}^{\mathrm{T}}$. In the first two cases, $\widetilde{\mathbf{A}}(d)$ is diagonal for any full rank leadfield matrix **L**, while in the third case it is still possible that $\widetilde{\mathbf{A}}(d)$ is non-diagonal depending on **L** (see appendix A for more details). Thus, non-interacting sources with proportional power spectra lead to non-interacting sensors, provided that the sensors are not more than the sources and are probed without measurement noise. Note that, in the particular case in which there is only one source ($M$=1) and the mixing still occurs without measurement noise, the sensor time series will be scaled versions of the source time series (with different scales determined by the entries of **L**). In such a case, Granger causality among the sensor time series is still absent because the inclusion of the other channels' past values would be completely redundant compared to the use of the target past; however, from a practical perspective, due to numerical inaccuracies and differences in algorithms, estimates of $\widetilde{\mathbf{A}}(d)$ will be highly unstable, yielding very different solutions or no-solution depending on the algorithm.

*3.2 Simulation I: Theoretical Connectivity Analysis of Uncoupled Sources*

In this section we report the analysis of GC in both time and frequency domains for a simple theoretical example demonstrating how spurious connectivity can be measured in the sensor space as a result of volume conduction. Specifically, we consider the case of two non-interacting sources described by the following MVAR process of order 2:

$$\begin{aligned} S_1(t) &= 0.95 S_1(t-1) - 0.70 S_1(t-2) + W_1(t) \\ S_2(t) &= 0.50 S_2(t-1) - 0.90 S_2(t-2) + W_2(t) \end{aligned} \tag{11}$$

where the innovations $W_1(t)$ and $W_2(t)$ are independently drawn from the Gaussian distribution with zero mean and unit variance. Then, mixed time series representative of the sensor activity are obtained by pre-multiplying $S(t) = [S_1(t)\ S_2(t)]^T$ with the following mixing matrix

$$\mathbf{L} = \begin{bmatrix} 0.8 & 0.3 \\ 0.4 & 0.7 \end{bmatrix}$$

and considering zero measurement noise, i.e., $Y(t) = \mathbf{L}\, S(t)$.

Then, the derivations in Sect 2.3 are exploited to define the SS model descriptive of this parameter setting, and connectivity analysis is performed in both time and frequency domains computing the TGC and DTF measures from the SS representation. The TGC is zero along both directions between the two sources $S_1$ and $S_2$ ($TGC_{12}$=$TGC_{21}$=0), while the results are substantially larger than zero between the two sensors $Y_1$ and $Y_2$ ($TGC_{12}$=0.1064, $TGC_{21}$=0.1417). The profiles of the DTF obtained between the two original and mixed sources are shown in Fig. 1 together with the corresponding spectral density functions. These results of both time and frequency domain analysis show no interactions among the original source time series. On the other hand, the mixed sources reveal a clear bidirectional interaction detected by significant TGC and DTF values along the two directions of interaction.

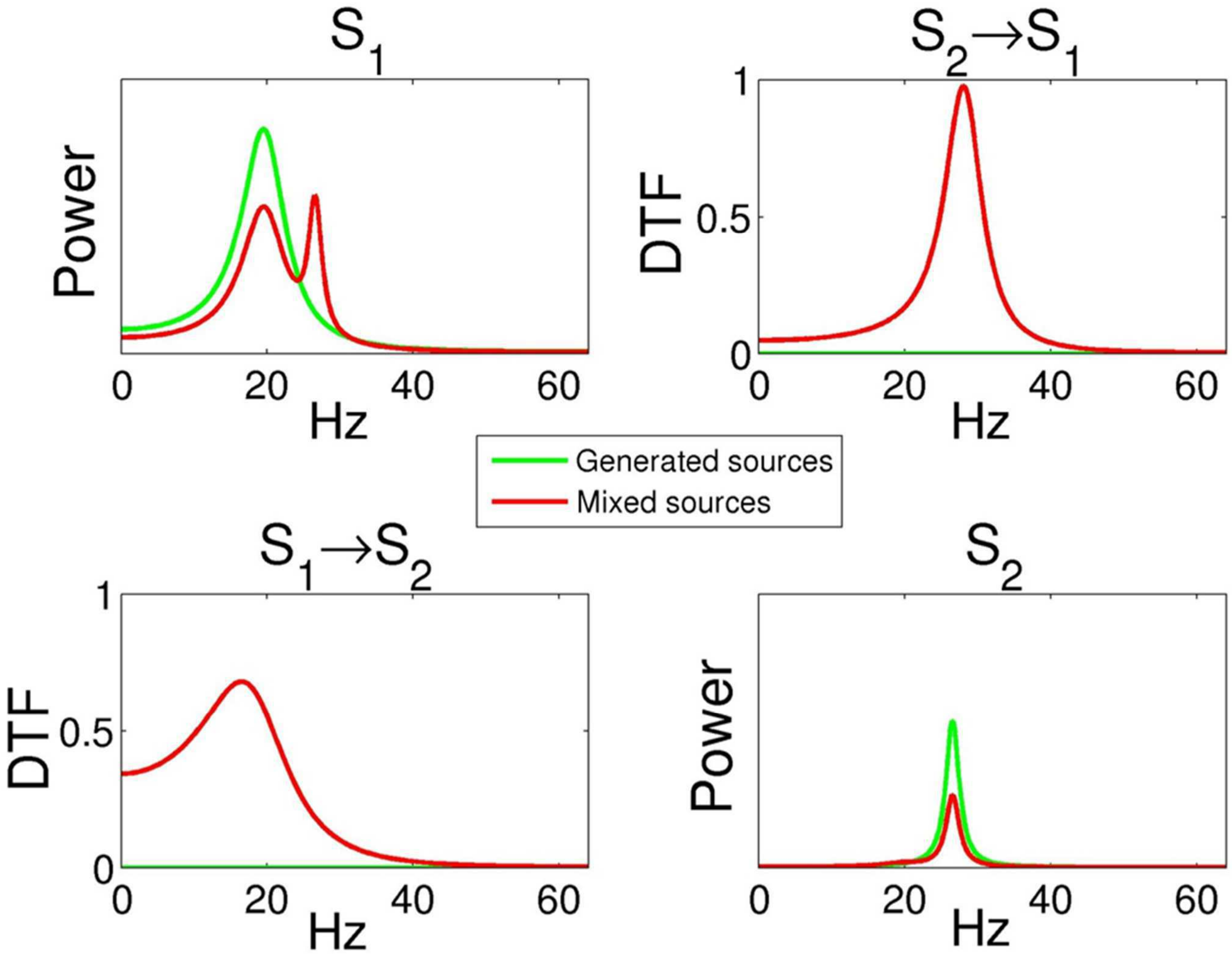

**Fig. 1** DTF analysis for simulation I, showing the profiles of DTF (off diagonal plots) and of the power spectrum (diagonal plots) of two non-interacting VAR sources before (green) and after (red) linear mixing. The simulated sampling frequency is 128 Hz

*3.2 Simulation II: Connectivity Analysis of a Realistic Pseudo-EEG dataset*

This second simulation was aimed at showing the effects of volume conduction on realistic pseudo-EEG data generated using an adapted version of the simulation framework proposed by Haufe and Ewald (2016). Here, we generated three sources with only one unidirectional coupling between two sources. The MVAR model of the sources was:

$$S_1(t) = 0.55S_1(t-1) - 0.80S_1(t-2) + W_1(t)$$
$$S_2(t) = 0.7S_2(t-1) - 0.5S_2(t-2) + 0.60S_1(t-1) + W_2(t)\,, \tag{12}$$
$$S_3(t) = 1.1S_3(t-1) - 0.50S_3(t-2) + W_3(t)$$

with noise terms drawn again from the Gaussian distribution with zero mean and unit variance. We see that the first source ($S_1$) Granger-causes the second source ($S_2$). The third source ($S_3$) is not connected to the first two sources.

First, we generated a realization of the three sources lasting $T$=10000 points, and computed the TGC and DTF over this realization from the MVAR parameters estimated according to the standard least squares approach (Faes et al. 2012). While the DTF was obtained straightforwardly as described in Sect. 2.2 from the MVAR parameters, the TGC was computed from the SS representation of the estimated MVAR parameters, thus avoiding the fit of full and reduced MVAR models (Barnett and Seth 2015). The TGC values were: $TGC_{12}$=$TGC_{13}$= $TGC_{23}$ = $TGC_{31}$ = $TGC_{32}$ = 0 and $TGC_{21}$=0.499. The DTF profiles are shown in Fig. 2.

Next, each of the three simulated sources was assigned to a specific anatomical location: Source one was assigned to the right occipital cortex (vertex MNI coordinates: [29 -89 4]), source two was assigned to the right middle orbital frontal cortex (vertex MNI coordinates: [41 52 -8], and the third source was assigned to the right superior temporal cortex (vertex MNI coordinates: [45 -15 3]), see fig. 3.

In addition to these three sources, 500 mutually statistically independent pink noise sources (i.e. brain noise) were generated and each brain noise source was randomly allocated to 1 of the vertices of the source model. All the generated sources were assumed to be dipoles perpendicularly oriented toward the surface. Here the ICBM-NY head combined with the boundary element method (BEM) was used to solve the forward problem (i.e. projecting source space time series to channel space time series using the lead field matrix). The sources were projected to 108 EEG channels defined in 10-5 electrode placement system (Oostenveld and Praamstra 2001). We used a similar procedure as in Haufe and Ewald (2016) for the projection of the brain sources (i.e. the MVAR sources and the 500 pink noise sources). More specifically, the MVAR sources and the brain noise sources were projected to the sensors, normalized by their Frobenius norm and summed:

$$Y_{brain}(t) = \mathrm{a}\frac{S(t)\boldsymbol{L}_{\boldsymbol{sources}}}{\|S(t)\boldsymbol{L}_{\boldsymbol{sources}}\|_F} + (1-\mathrm{a})\frac{Z(t)\boldsymbol{L}_{bnoise}}{\|Z(t)\boldsymbol{L}_{bnoise}\|_F} \tag{13}$$

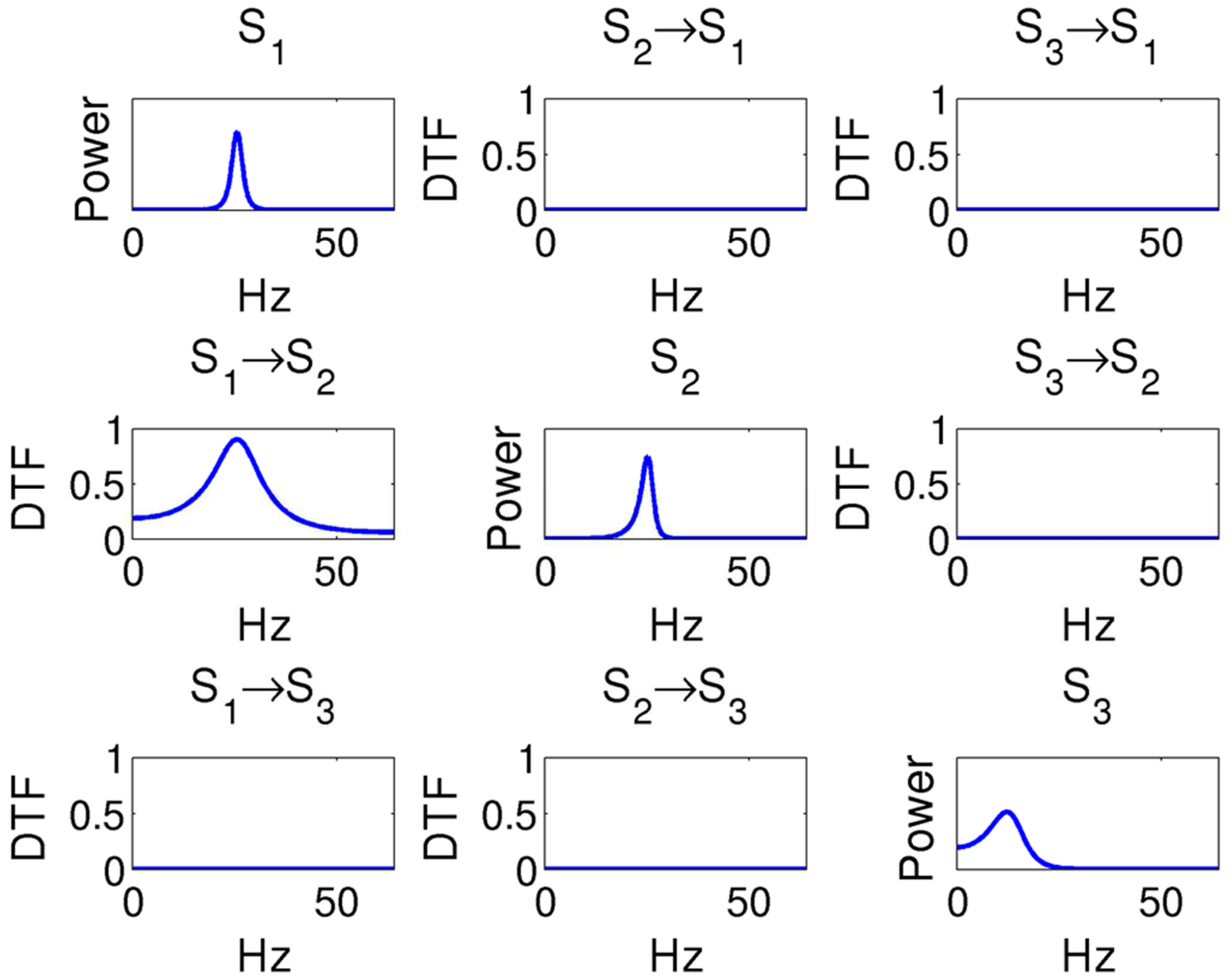

**Fig. 2** The DTF for the generated sources are show. The diagonal panels show the power spectrum of the three sources

where $Y_{brain}$(t) are the channel time series without measurement noise. *S*(t) and *Z*(t) are the three MVAR sources and the 500 brain noise sources respectively.$||Q(t)||_F$ denotes the Frobenius norm of *Q(t)*. The matrices $\boldsymbol{L_{sources}}$ and $\boldsymbol{L_{bnoise}}$ are the leadfield matrices for the MVAR sources and the brain noise respectively. The signal-to-noise (SNR) parameter ,a, was set to 0.8. The overall channel data, *Y*(t), was generated according to

$$Y(t) = 0.9\,\frac{Y_{brain}(t)}{\|Y_{brain}(t)\|_F} + 0.1\,\frac{G(t)}{\|G(t)\|_F} \tag{14}$$

where *G*(t) is independent measurement noise drawn from the standard normal distribution.

The following channels were selected for connectivity analysis in the sensor space: AF8,O2 and T8.

These channels were selected as those being the closest to the three source locations. We applied eLORETA to the channel time series to reconstruct the three MVAR sources at the *a priori* defined anatomical locations (Pascual-Marqui 2007). An MVAR model of order 2 was fitted to these reconstructed time series using ordinary least squares method. The DTF and TGC were derived from the estimated MVAR parameters. In addition to the two stage approach, we applied the expectation maximization algorithm developed by (Cheung et al. 2010) to estimate the SS parameters. This algorithm was developed specifically for EEG (MEG) applications and requires that the source locations to be specified *a priori* (see Cheung et al. 2010 for more details).The DTF and TGC were obtained from the estimated state equation parameters. The *TGC* values and DTF profile for the sensor-space analysis can be found in table 1 and fig. 4 respectively. The TGC values and DTF profiles for source reconstructed analysis and SS analysis can be found in table 2 and fig. 5. The sensor-space DTF analysis reveals clear non-zero DTF's from O2 to AF8, from O2 to T8 and from AF8 to T8. The DTF for the other connections are much smaller. The trends of DTF for the source-reconstructed time series and SS analysis were almost identical to those of the originally generated time-series.

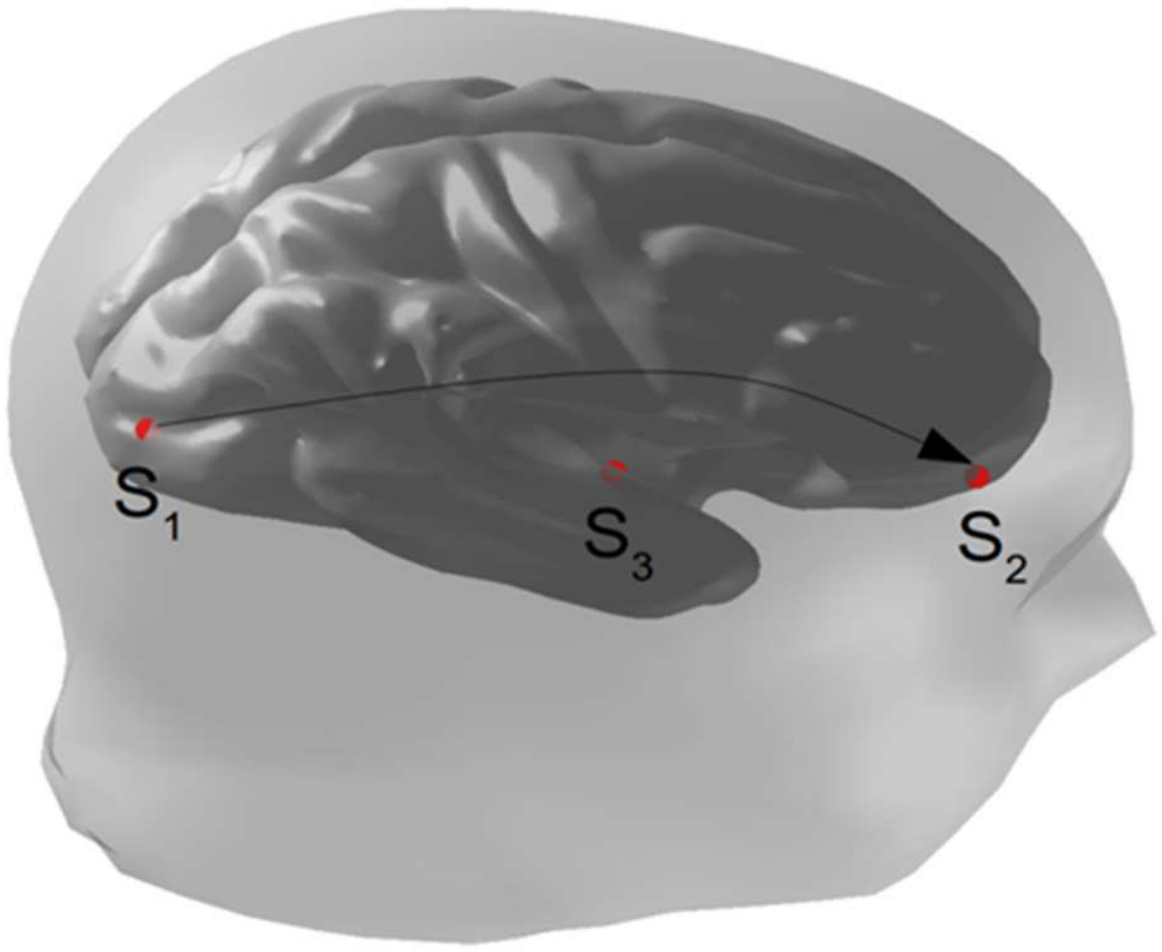

**Fig. 3** Anatomical locations of the three generated sources (red dots). The arrow indicates the unidirectional connection from source one to source two

**Table 1** Values of sensor-space *TGC*

| | Driver channel | | |
|---|---|---|---|
| Target channel | O2 | AF8 | T8 |
| O2 | | 0.0199 | 0.0042 |
| AF8 | 0.1120 | | 0.0807 |
| T8 | 0.0763 | 0.0123 | |

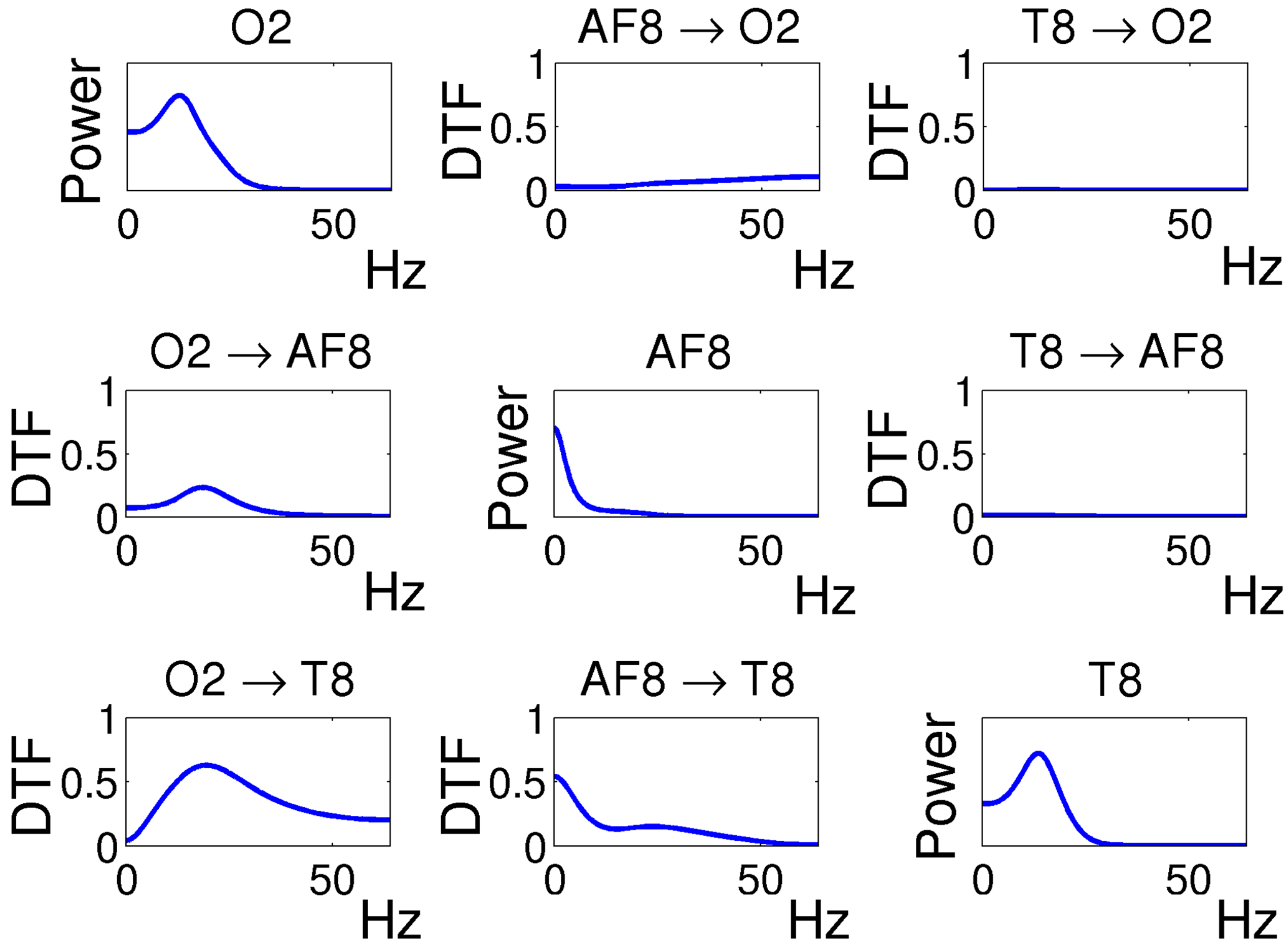

**Fig. 4** Results of causal connectivity analysis showing DTF and power spectra (on the diagonal) computed for sensor space analysis

**Table 2** Values of source-reconstructed and SS-based *TGC*

| SS analysis | | | | Source reconstructed analysis | | | |
|---|---|---|---|---|---|---|---|
| | Driver source | | | | Driver source | | |
| Target source | $S_1$ | $S_2$ | $S_3$ | Target source | $S_1$ | $S_2$ | $S_3$ |
| $S_1$ | | 0.0011 | 0.0002 | $S_1$ | | 0.0028 | 0.0000 |
| $S_2$ | 0.4938 | | 0.0009 | $S_2$ | 0.4132 | | 0.0008 |
| $S_3$ | 0.0006 | 0.0004 | | $S_3$ | 0.0030 | 0.0037 | |

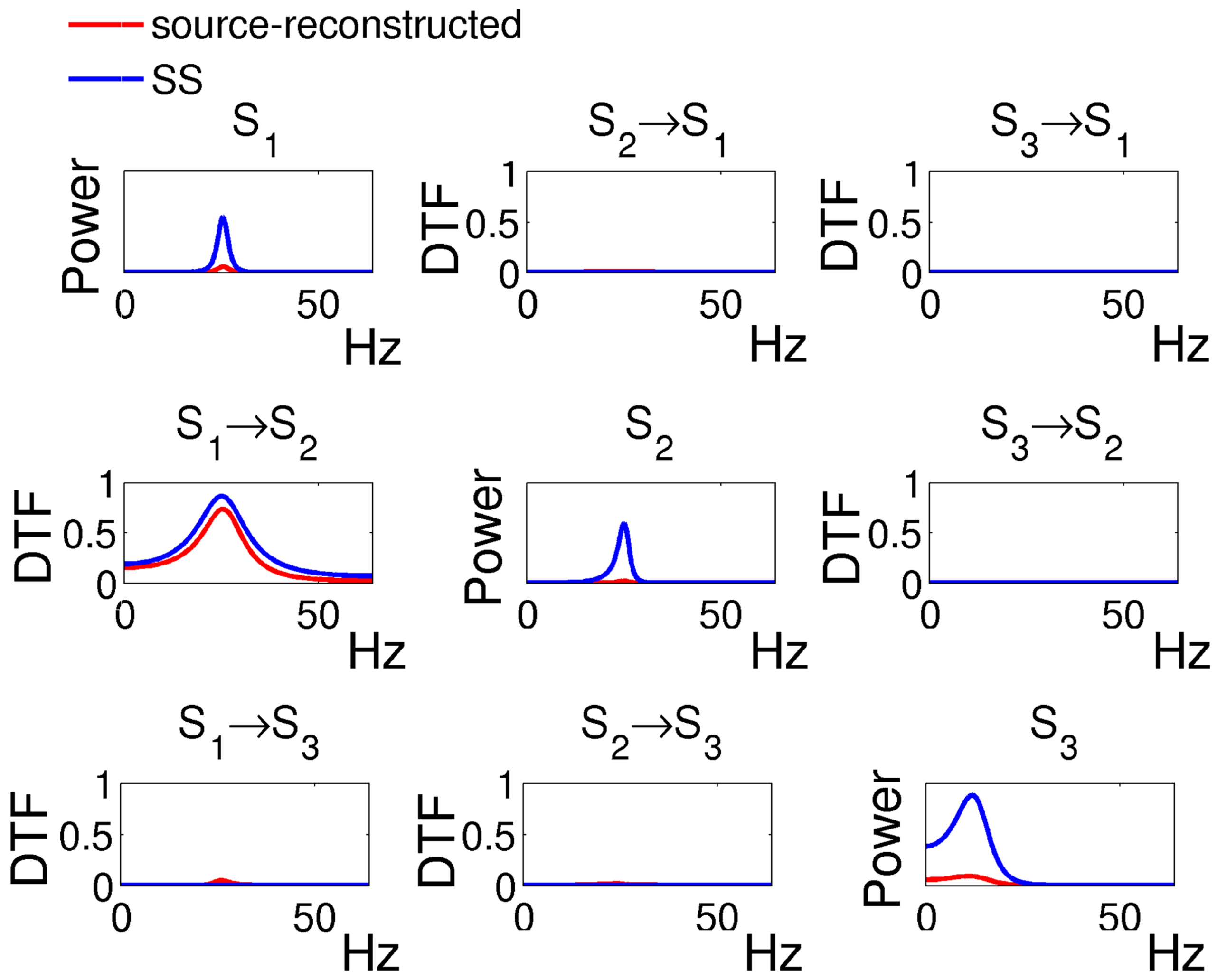

**Fig. 5** Results of causal connectivity analysis showing DTF and power spectra (on the diagonal) computed for source-reconstructed and SS analysis

## 4. Discussion

The main objective of this study was to show that the use of traditional MVAR-based connectivity analysis, yielding classic measures such as the GC in the time domain and the DTF in the frequency domain, does not allow interpretation of anatomically interacting brain areas when performed on EEG sensor-space. First of all, due to volume conduction there is no one-to-one relation between the EEG channel time series and source activity. More specifically, multiple spatially distinct neural sources contribute to the measured EEG scalp potentials. In addition, the depth and orientation of the sources has a major impact on how a source is projected to the channels. A single deep source for example will have a very diffuse effect across many channels. A tangential source that is relatively close to the surface will result in larger activity in channels that are not closest to that source location. A perpendicular source close to the surface will project most strongly to the nearest channel. See fig 6 for an instructive example of the effect of the orientation of a source that is close to the surface.

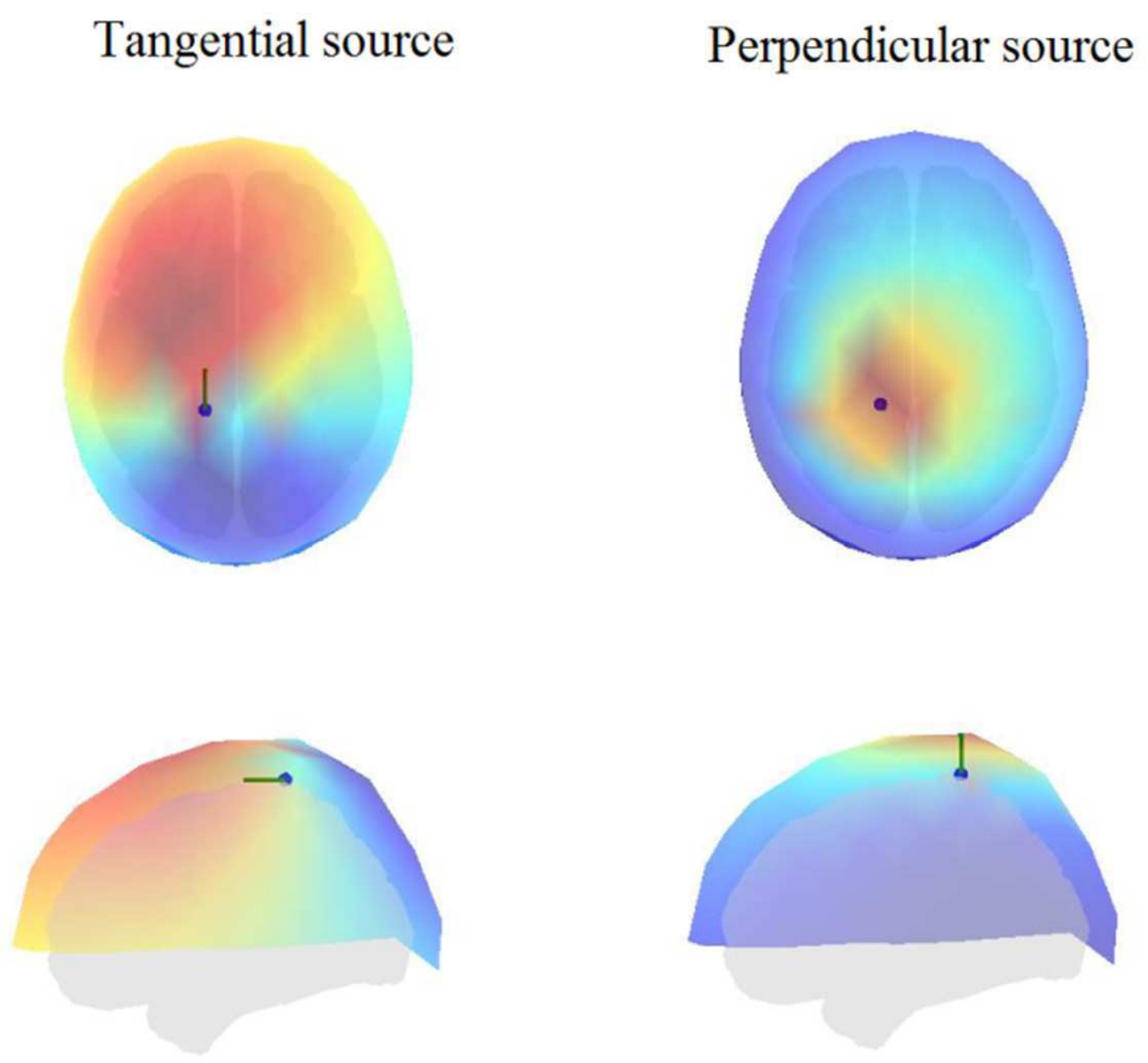

**Fig. 6** An instructive example of the effect of source orientation on the scalp electrodes. The sources are of unit strength and projected to the scalp. Top row and bottom row shows the locations (dot), orientation (arrow) of the sources (left: tangential source; right: perpendicular source) and corresponding topography in axial and sagittal view respectively

Therefore the spatial location of the EEG channels cannot be seen as an approximation of a source's anatomical location (such as the anatomical location underneath the electrode). Secondly, we show that, since the EEG time series are a mixture of sources, spurious connectivity can occur quite easily between pairs of channels even in the absence of any causal interaction at the source level. Since the latter has been debated in the past by Kaminski and Blinowska (2014), this issue is the main focus of this paper. More specifically, the authors claim that the DTF is not influenced by volume conduction. Volume conduction in EEG is mathematically equivalent to a linear superposition (mixing) of brain sources. If the time evolution of one or more sources can be characterized by a VAR model, then the mixed time series will also contain an autoregressive component. However, since the mixing of sources is not identical for the different channels, the inclusion of past values of another (driver) channel can improve the prediction of another (target) channels' current state. The same argument was given in (Haufe et al. 2013). In that paper, the authors state that the even in the case only one MVAR source is active, the inclusion of some channels' past values will improve the prediction of the current state in another channel because both channels measure the same source but with different noise realizations (i.e. measurement noise). Crucial in their argument is the measurement noise. In case of no measurement noise, the two channels will be differently scaled versions of that source. as we also discuss in Sect. 3.1, no causal connectivity will be present according to the Granger's principle. In addition, the adverse effect of additive noise (and more specifically linearly mixed additive noise, which can be caused by volume conduction) on GC-based connectivity analysis was also shown by Vinck et al. (2015)

In case of *multiple* non-interacting MVAR sources, differentially mixing of these sources will also result in spurious connections even in the absence of additive measurement noise. In this case the inclusion of another channel past values in the model will not introduce redundancy. This was shown in the first simulation where two non-interacting MVAR sources were specified and then differentially mixed to obtain two mixed time-series. No measurement noise was added to the mixed time series. The DTF was calculated on both the original and the mixed time series. In fig. 1, one can clearly see that, although the DTF for the original sources is uniformly zero over frequencies in both directions, the DTF of the mixed times-series clearly show a bidirectional interaction. Kaminski and Blinowska (2014), however, claim that the DTF is insensitive to volume conduction. They argue that the linear superposition of the sources is instantaneous and hence cannot produce a phase difference at the sensors. However, linear superposition of sources can change the phase difference at the sensors under certain conditions (see appendix B for an illustration).In the case of non-interacting sources, however, the phase of the cross spectrum will be zero. Nevertheless, it is not true that signals without a phase difference should result in vanishing DTF. An example of this is given by our first simulation (Eq. (11) and Fig. 1), for which the cross-spectrum computed either between two sources or between the two sensors has vanishing phase, while the DTF is zero if computed between the sources and nonzero if computed between the sensors. Here, the motivation behind the non-vanishing sensor DTF is not the presence or absence of phase difference, but rather the mixing of the same source activity in the sensor space, which lets the source act as a 'common driver' of the activity measured at the sensors; the result is that, given that these common effects are not (and cannot be) conditioned out in causal connectivity analyses, the time lagged dependencies within the source are misinterpreted as time-lagged cross-dependencies between the sensors into which these effects are mixed, even though the mixing occurs instantaneously.

The bottom line is that volume conduction can result in a change in phase difference under certain conditions and that a vanishing phase difference does not imply a vanishing DTF. Of course, the same holds for many other measure of causal connectivity computed in the time domain (e.g., the TGC studied here) or in the frequency domain (e.g., the partial directed coherence (Baccala et al. 2001)). On the contrary, in this study we have shown that the measures of causal connectivity are in general sensitive to volume conduction, while they are robust to it –in the sense that they do not indicate false positive sensor connectivity in the absence of source connectivity– only in some particular conditions that include the lack of autocorrelations for each individual source, the presence of the same autocorrelation structure for all sources and in the case of a single MVAR source that is projected without measurement noise. These conditions are arguably unrealistic in case of real EEG data. First of all, the absence of uncorrelated measurement noise is highly unlikely (if not impossible). Secondly, even if the (few) sources of interest show the same autocorrelation structure, many other brain sources that are not of interest (i.e. brain noise sources) will be superimposed on the sensor level. It is however unreasonable that all of these brain noise sources are either none interacting ,show no autocorrelation or have the same autocorrelation structure as the signals of interest. Therefore, sensor space analysis will in general show significant connectivity even if the sources of interest are non-interacting.

Furthermore, Kaminski and Blinowska (2014) suggest that no preprocessing of the channels should be done to reduce the volume conduction effect because these procedures would destroy the causal information between sources. In our second simulation study, we show that (in principle) applying the DTF to source reconstructed time-series or by means of the SS approach can result in the correct recovery of the original (anatomical) causal structure of the dynamical system and is thus not a priori wrong. It should be noted that in our simulations, we used the same head model for source reconstruction and the time series were obtained at the a priori spatial locations. In practice, however, the head model can only be an approximation of the 'true' head model. It has been shown that errors in the head model can result in errors in source-space GC based connectivity analysis (Cho et al. 2015). In addition, the true locations of activated sources are not known a priori and hence need to be chosen based on the data itself or prior knowledge (e.g. from fMRI studies using a similar paradigm; see e.g. Schoffelen and Gross (2009) for selection strategies of sources in the context of source connectivity analysis). Since source-space causal connectivity analysis requires estimates of the source time series, a source reconstruction algorithm needs to be applied to the data. Since the inverse problem is ill-posed, the solution depends on (biologically) informed constraints. Different inverse solutions pose different constraints and the accuracy of the output of the algorithm depends on the correctness of these constraints. Furthermore, inverse solution does not undo the mixing completely and hence spatial leakage will still be present. This is especially the case for sources that are close to each other. In our simulations, the sources were placed far apart from each other so that mixing effects at these source locations was minimal. Nevertheless, if we would have taken a source estimate close to one of the two interacting sources, it is likely that we would have found spurious connectivity. We emphasize that when applying source reconstruction on real EEG data, mixing effects in the source space remain. Nevertheless, we show that use of the DTF on source reconstructed time series is not a priori wrong and may be used under certain circumstances.

In order to overcome the leakage effects in the source space, other methods that are more robust to mixing are required to assess brain interaction between the estimated source time series. One such method that is specifically designed for measures such as TGC and DTF is time reversal (Haufe et al. 2013). Time reversed data are used as surrogates for statistical testing, effectively alleviating spurious causality due to volume conduction Importantly, (under certain conditions) reversal testing correctly identifies causality in the presence of true interaction (Winkler et al. 2016). Although the methods that are robust to mixing can be applied to the sensor space, their anatomical interpretability is still problematic given that the sensor location cannot be seen as proxy to anatomical location. Moreover, (Mahjoory et al. 2016) have shown that the choice of reference has a major impact on the sensor connectivity profile. Using the phase slope index (i.e. a robust measure that assess the directionality of information flow, see Nolte et al. 2008), they showed that changing the reference can reverse the directionality of the connectivity pattern. In sum, even though source-space causal connectivity analysis is a promising avenue, many issues still needs to be addressed in future research.

Another possibility is to use the SS framework to assess causal interactions among neural sources. When using SS models for EEG, both volume conduction and the temporal evolution of the neural state are modeled within the same framework. The SS framework has, for example, been used to solve the dynamic inverse solution in EEG, which results in better spatiotemporal recovery of the hidden neural sources (Galka et al. 2004; Yamashita et al. 2004). In the context of causal connectivity analysis, Cheung et al. (2010), recently developed an expectation maximization algorithm to obtain model parameter of (8a) and (8b). In that algorithm, the anatomical locations of the sources needs to be specified *a priori* and hence prior knowledge or EEG source imaging needs to be applied first. Nevertheless, the authors showed that GC analysis using SS models was less sensitive to noise compared to the two-stage procedure. In addition, other issues (such as downsampling and other preprocessing procedures) with respect to the MVAR modeling in causal connectivity analysis can be overcome by using the SS framework (see Barnett and Seth 2015). In the second simulation, we showed that the connectivity analysis using the SS framework resulted also in the correct recovery of the DTF profiles.

To conclude, sensor-space causal connectivity analysis, performed either in time or frequency domain, does not allow interpretation in terms of interacting brain sources. We also showed that the use of measures such as the DTF and TGC in the source space is not a priori wrong as claimed by Kaminski and Blinowska (2014). However, in order to mitigate mixing effect in the source space, it is advised to used measures which are robust to volume conduction.

Appendix A

The sources can be written as

$$S(t) = \boldsymbol{L}^{\dagger}\big(Y(t) - G(t)\big) + Z(t). \tag{A1}$$

Assuming that **L** is full rank and Y(t) is in the range of **L**, (A1) has at least one solution where *Z*(t) is in the nullspace of **L**. **L**† denotes the pseudo inverse of L. Substituting (A1) into (10a) gives

$$Y(t) = \sum_{d=1}^{p} \mathbf{LA}(d)\big\{\boldsymbol{L}^{\dagger}\big(Y(t-d) - G(t-d)\big) + Z(t-d)\big\} + G(t) + \boldsymbol{L}U(t) \tag{A2a}$$

$$Y(t) = \sum_{d=1}^{p} \mathbf{LA}(d)\boldsymbol{L}^{\dagger}Y(t-d) + \sum_{d=1}^{p} \mathbf{LA}(d)Z(t-d) - \sum_{d=1}^{p} \mathbf{LA}\boldsymbol{L}^{\dagger}(d)G(t-d) + G(t) + \boldsymbol{L}U(t) \tag{A2b}$$

We can define

$$\widetilde{\boldsymbol{A}}(d) = \boldsymbol{LA}(d)\boldsymbol{L}^{\dagger} \text{ and } \widetilde{U}(t) = \boldsymbol{L}u(t). \tag{A3}$$

Assuming no measurement noise, (A2b) becomes

$$Y(t) = \sum_{d=1}^{p} \widetilde{\boldsymbol{A}}(d)Y(t-d) + \sum_{d=1}^{p} \mathbf{LA}(d)Z(t-d) + \widetilde{U}(t). \tag{A4}$$

Suppose $\boldsymbol{A}_d = a_d\boldsymbol{I}$ for all d. We can describe three cases: L is square, fat or skinny.
When *L is square*, we have that *Z*(t) =0 and **L**†=$\mathbf{L}^{-1}$. Hence

$$\widetilde{\boldsymbol{A}}(d) = \boldsymbol{LA}(d)\boldsymbol{L}^{\dagger} = a_d\boldsymbol{LL}^{-1} = a_d\boldsymbol{I}. \tag{A5}$$

When *L is fat*, **LA**(d)*Z*(t-d)= $a_d$**L***Z*(t-d)=0 and $\boldsymbol{L}^{\dagger} = \boldsymbol{L}^{T}(\boldsymbol{LL}^{T})^{-\mathbf{1}}$, hence

$$\widetilde{\boldsymbol{A}}(d) = \boldsymbol{LA}(d)\boldsymbol{L}^{\dagger} = a_d\boldsymbol{LL}^{T}(\boldsymbol{LL}^{T})^{-\mathbf{1}} = a_d\boldsymbol{I}. \tag{A6}$$

When *L is skinny* we have that *Z*(t) =0 and $\boldsymbol{L}^{\dagger} = (\boldsymbol{L}^{T}\boldsymbol{L})^{-\mathbf{1}}\boldsymbol{L}^{T}$. Hence

$$\widetilde{\boldsymbol{A}}(d) = \boldsymbol{LA}(d)\boldsymbol{L}^{\dagger} = a_d\boldsymbol{L}(\boldsymbol{L}^{T}\boldsymbol{L})^{-\mathbf{1}}\boldsymbol{L}^{T}. \tag{A7}$$

From the derivations above we can conclude that in the case of no measurement noise and when $\boldsymbol{A}_d = a_d\boldsymbol{I}$ for all d, non-interacting sources leads to non-interacting sensors when **L** is either fat or square. When **L** is skinny, spurious interaction at the sensors can occur.

# Appendix B

Here phase difference is defined as the phase of the cross spectrum. The cross-spectrum of two time series is (the frequency index is drop for notational simplicity):

$$C_{S_1S_2} = E\{S_1S_2^*\} \tag{B1}$$

Where the phase of the cross spectrum can be obtained by:

$$\phi_{S_1S_2} = \arctan\left(\frac{\Im(C_{S_1S_2})}{\Re(C_{S_1S_2})}\right) \tag{B2}$$

Suppose the time series $S_1$ and $S_2$ are linearly superimposed into time series $Y_1$ and $Y_2$ by:

$$\begin{bmatrix} l_{11} & l_{12} \\ l_{21} & l_{22} \end{bmatrix} \begin{bmatrix} S_1 \\ S_2 \end{bmatrix} = \begin{bmatrix} Y_1 \\ Y_2 \end{bmatrix} \tag{B3}$$

Then the cross-spectrum of $Y_1$ and $Y_2$ equals

$$C_{Y_1Y_2} = E\{(l_{11}S_1 + (l_{12}S_2)(l_{21}S_1 + (l_{22}S_2)^*\} \tag{B4a}$$

$$C_{Y_1Y_2} = l_{11}l_{21}E\{S_1S_1^*\} + l_{12}l_{22}E\{S_2S_2^*\} + (l_{11}l_{22} - l_{12}l_{21})E\{S_1S_2^*\} \tag{B4b}$$

The phase difference between $Y_1$ and $Y_2$ can then be formulated as

$$\phi_{Y_1Y_2} = \arctan\left(\frac{(l_{11}l_{22} - l_{12}l_{21})E\{\Im(S_1S_2^*)\}}{l_{11}l_{21}E\{|S_1|^2\} + l_{12}l_{22}E\{|S_2|^2\} + (L_{11}L_{22} - L_{12}L_{21})E\{\Re(S_1S_2^*)\}}\right) \tag{B5}$$

From (B5), we can derive two conditions in which the linear superposition does not change the phase difference (i.e. when (B5) = (B2)): 1) when L is either diagonal or anti diagonal and 2) when (16) is zero or π (i.e when the sources have a phase difference of 0 or π and hence the imaginary part of their cross spectrum is zero). To illustrate that volume conduction can lead to a phase shift. Two signals were generated with a phase difference of π/3. The off diagonal of the leadfield matrix was varied according to $L_{12}= L_{11}k$ and $L_{21}= L_{22}k$.The phase of the cross spectrum was then calculated for k =0,..2 See fig. B1 for the results.

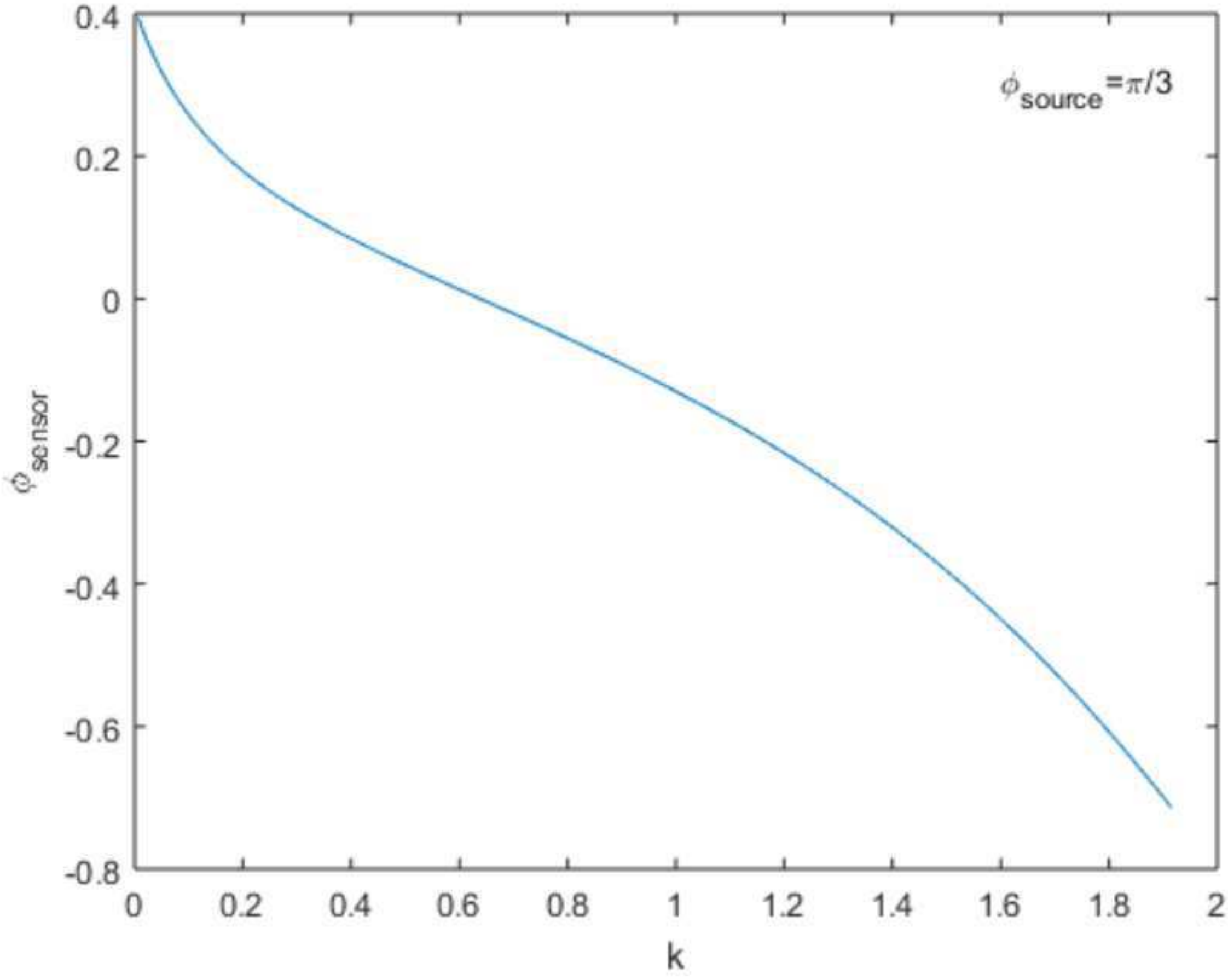

**Fig. B1** The phase difference of the mixed signals is plotted against varying values of k.

To conclude under certain conditions, volume conduction can lead to a change in phase difference